\documentclass[lettersize,journal]{IEEEtran}
\usepackage{amsmath,amsfonts}
\usepackage{algorithmic}
\usepackage{algorithm}
\usepackage{array}
\usepackage[caption=false,font=normalsize,labelfont=sf,textfont=sf]{subfig}
\usepackage{textcomp}
\usepackage{stfloats}
\usepackage{url}
\usepackage{verbatim}
\usepackage{graphicx}
\usepackage{xcolor}
\usepackage{cite}
\usepackage{multirow} 
\usepackage{booktabs} 

\begin{document}

\title{LoD-Structured 3D Gaussian Splatting for \\Streaming Video Reconstruction}

\author{Xinhui Liu, Can Wang, Lei Liu, Zhenghao Chen, Wei Jiang, Wei Wang, Dong Xu, \IEEEmembership{Fellow, IEEE} 
\thanks{X. Liu, C. Wang and L. Liu are with the School of Computing and Data Science, The University of Hong Kong, Hong Kong, China. E-mail: (xhliu01, canwang, and liulei95)@hku.hk.}
\thanks{W. Jiang and W. Wang are with the Futurewei Technologies Inc, Santa Clara, CA, USA. E-mail: (wjiang, rickweiwang)@futurewei.com. }
\thanks{Z. Chen is with the School of Computer and Information Sciences, The University of Newcastle, NSW, AUS. E-mail: zhenghao.chen@newcastle.edu.au.}
\thanks{D. Xu is with the School of Computing and Data Science, The University of Hong Kong, Hong Kong, China. E-mail: dongxu@hku.hk. Corresponding Author.}
}

\markboth{Journal of \LaTeX\ Class Files,~Vol.~14, No.~8, August~2021}%
{Shell \MakeLowercase{\textit{et al.}}: A Sample Article Using IEEEtran.cls for IEEE Journals}

\maketitle

\begin{figure*}[!t]
\centering
\includegraphics[width=7.1in]{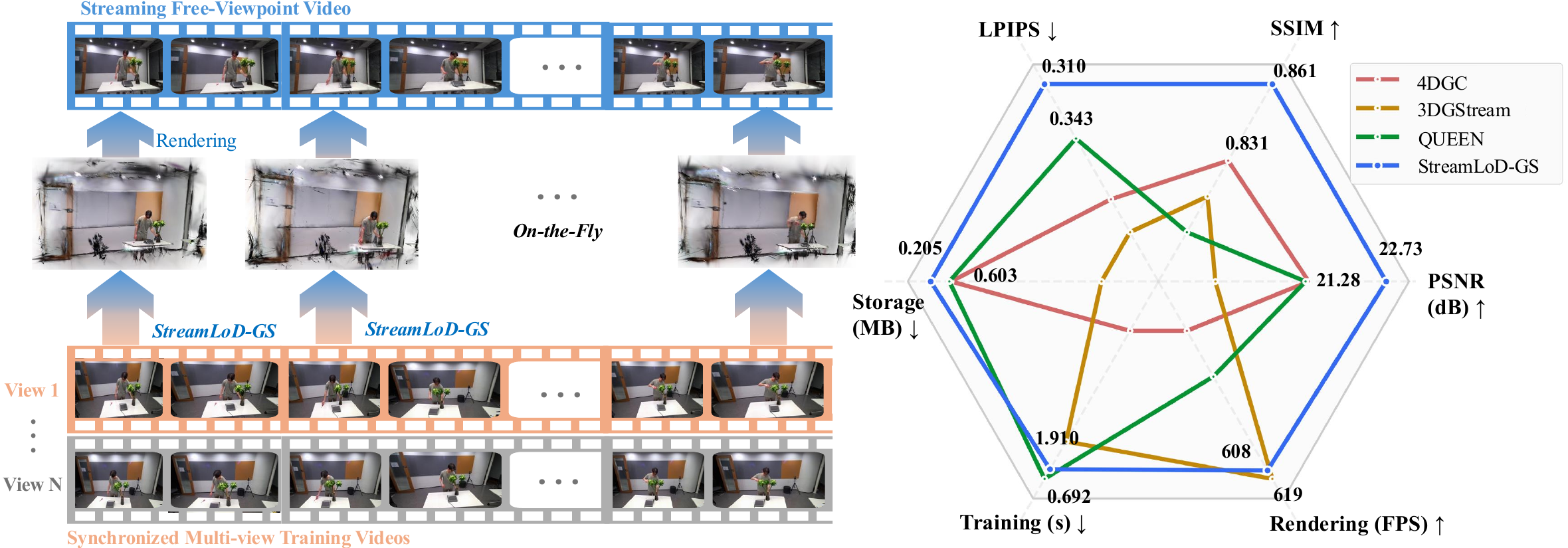}
\caption{StreamLoD-GS enables fast, on-the-fly, high-fidelity reconstruction for Streaming Free-Viewpoint Video. \textbf{Right}: Comparison with existing methods with 5 Training Views on Meet Room~\cite{li2022streaming} dataset.}
\label{fig:intro}
\end{figure*}

\begin{abstract}
Free-Viewpoint Video (FVV) reconstruction enables photorealistic and interactive 3D scene visualization; however, real-time streaming is often bottlenecked by sparse-view inputs, prohibitive training costs, and bandwidth constraints.
While recent 3D Gaussian Splatting (3DGS) has advanced FVV due to its superior rendering speed, Streaming Free-Viewpoint Video (SFVV) introduces additional demands for rapid optimization, high-fidelity reconstruction under sparse constraints, and minimal storage footprints.
To bridge this gap, we propose StreamLoD-GS, an LoD-based Gaussian Splatting framework designed specifically for SFVV. Our approach integrates three core innovations: 1) an Anchor- and Octree-based LoD-structured 3DGS with a hierarchical Gaussian dropout technique to ensure efficient and stable optimization while maintaining high-quality rendering; 2) a GMM-based motion partitioning mechanism that separates dynamic and static content, refining dynamic regions while preserving background stability; and 3) a quantized residual refinement framework that significantly reduces storage requirements without compromising visual fidelity. 
Extensive experiments demonstrate that StreamLoD-GS achieves competitive or state-of-the-art performance in terms of quality, efficiency, and storage. 
\end{abstract}
\begin{IEEEkeywords}
3DGS, Sparse-view, Free-Viewpoint Videos, Video Streaming.
\end{IEEEkeywords}

\section{Introduction}
\IEEEPARstart{C}{onstructing} Free-Viewpoint Videos (FVVs) from multi-view images promise to revolutionize visual media by bridging the gap between our perception of the dynamic 3D world and the constraints of conventional 2D video~\cite{chen2019overview}. 
Recently, 3D Gaussian Splatting (3DGS)~\cite{kerbl20233d} has emerged as a cornerstone for FVV reconstruction~\cite{collet2015high,chen2025livegs,chen2019overview,zhang2025mega,tang2025compressing,kim20244d,wu20244d}, primarily due to its capacity for photorealistic and real-time synthesis.
Due to the increasing demand for real-time interactivity and immersive experiences, streaming Free-Viewpoint Video (SFVV)~\cite{sun20243dgstream,gao2024hicom,girish2024queen,yan2025instant,han2024event,hu20254dgc} has emerged as a focal point of research, offering dynamic, real-time viewing experiences that transcend traditional media boundaries.  
However, SFVV introduces new challenges and demands for 3DGS, including: 1) Fast optimization and rendering: the ability to rapidly optimization and render novel views in real-time; 2) High fidelity even with sparse views: ensuring the visual quality of rendered scenes, even when only limited camera setups are available, as capturing and transmitting dense views in real-world environments is often logistically and financially prohibitive; 3) low storage requirements for transport (bandwidth): minimizing the storage of 3DGS points, to optimize bandwidth usage.  

To address these challenges, 3DGStream~\cite{sun20243dgstream} pioneered the extension of 3DGS to SFVV by leveraging Instant-NGP~\cite{muller2022instant}, which combines neural network representations and multi-resolution hashing to accelerate data querying and rendering. Beyond representation, other contemporary approaches~\cite{yan2025instant,girish2024queen,gao2024hicom,hu20254dgc} have shifted focus toward motion field modeling, utilizing residual learning~\cite{wang2023neural} to isolate dynamic regions and reduce optimization costs. 
Despite these advancements, most 3DGS-based frameworks~\cite{sun20243dgstream,gao2024hicom,girish2024queen,yan2025instant} still depend on dense, synchronized camera arrays to provide sufficient views for high-quality reconstruction. 
In practice, however, factors such as bandwidth constraints and spatial limitations often restrict systems to only a few cameras, hindering their applicability in real-world scenarios.
When attempting SFVV reconstruction under sparse training views, two primary challenges emerge.
\underline{First,} as the number of input views decreases ($e.g.$, 7, 5, 4, or fewer), weakened multi-view consistency causes the model to overfit to observed views, leading to artifacts in novel-view synthesis.
\underline{Second,} sparse inputs often trigger the production of redundant or misplaced Gaussians during the densification process~\cite{park2025dropgaussian,xu2025dropoutgs}, which inflates storage and transmission overhead that are particularly critical for mobile and low-end devices with limited memory and bandwidth.

Drawing inspiration from Level-of-Detail (LoD) hierarchies~\cite{luebke2002level,ren2025octree,shen2025lod,seo2024flod,kulhanek2025lodge}, which effectively reduce geometric redundancy while enhancing rendering efficiency, we propose StreamLoD-GS, an LoD-based Gaussian Splatting framework tailored for SFVV. Our approach comprises three core innovations:
\underline{First,} we introduce an LoD-structured 3DGS with Anchor and Octree representations. To prevent overfitting and facilitate stable optimization, we introduce a hierarchical Gaussian dropout technique. This strategy dynamically selects detail levels based on viewing distance and employs level-aware stochastic dropout during training. Ultimately, this design mitigates Sparse-View  overfitting while preserving the gradient flow essential for robust convergence.  
\underline{Second,} leveraging the inherent temporal coherence in video sequences, we develop a Gaussian Mixture Models~\cite{permuter2006study} (GMM)-based motion partitioning mechanism. This module intelligently identifies dynamic anchors for selective refinement while freezing static regions, thereby exploiting background stability and motion continuity to significantly reduce redundant computations.
\underline{Finally,} to enable bandwidth-efficient streaming, we incorporate a quantized residual refinement framework. This design bridges static frozen anchors and dynamically updated regions through quantized residuals, drastically minimizing storage and transmission footprints without sacrificing visual fidelity.

Our major contributions are summarized as follows.
\begin{itemize}
\item We propose an Anchor- and Octree-based LoD-structured 3DGS representation, integrated with a hierarchical Gaussian dropout technique to stabilize training and ensure high-fidelity synthesis in Sparse-View SFVV.
\item We introduce a GMM-driven mechanism to decouple dynamic and static content, enabling targeted refinement of moving regions while maintaining the structural stability of the background.
\item We develop a quantized residual refinement framework that compresses dynamic updates, facilitating efficient data transmission and storage for low-bandwidth environments. 
\item Extensive experiments demonstrate that StreamLoD-GS achieves competitive or state-of-the-art performance in terms of quality, efficiency, and storage footprints compared to contemporary SFVV approaches.
\end{itemize}

\section{Related Works}
\label{sec:relatedeork}

\subsection{Streaming 3DGS} 
Constructing streamable dynamic scenes via on-the-fly, frame-by-frame training presents significant challenges compared to offline learning from complete multi-view videos. Initial efforts adapted NeRFs~\cite{mildenhall2021nerf,barron2021mip,barron2022mip,cao2023hexplane,fridovich2023k,muller2022instant} to this task. For instance, StreamRF~\cite{li2022streaming} utilized an incremental learning paradigm by modeling per-frame differences, while ReRF~\cite{wang2023neural} modeled residual information between adjacent frames for long-duration scenes. NeRFPlayer~\cite{song2023nerfplayer}, on the other hand, decomposed the 4D spatio-temporal space to optimize for model compactness and reconstruction speed.
The advent of 3DGS marked a major leap forward, offering superior training efficiency and rendering speeds. Building on this, streaming 3DGS is designed to reconstruct dynamic scenes through on-the-fly, frame-by-frame training, enabling real-time updates and adjustments as the scene evolves~\cite{sun20243dgstream,gao2024hicom,yan2025instant,liu20253d,girish2024queen,hu20254dgc,tang2025compressing}. 
3DGStream~\cite{sun20243dgstream} proposed a Neural Transformation Cache with multi-resolution hashing to advance 3DGS modeling. 
Other methods have focused on advanced motion modeling, such as HiCoM~\cite{gao2024hicom} with its hierarchical coherent motion mechanism and IGS~\cite{yan2025instant}, which uses an anchor-driven network to learn motion residuals in a single step for faster training. 
To enhance model expressiveness and tackle data size, QUEEN~\cite{girish2024queen} proposed a framework based on Gaussian residuals combined with learned quantization and sparsity for compression. 
To optimize the rate-distortion trade-off for SFVV, 4DGC~\cite{hu20254dgc} introduces a rate-aware compression framework with motion-grid-based representation and differentiable quantization.  
Despite advancements in fidelity and efficiency, a common bottleneck remains: these methods require dense, multi-view images at every frame for 3D reconstruction. This dependence on synchronized multi-camera setups significantly limits their practical applicability. Furthermore, these approaches often necessitate a large number of Gaussian primitives to model the scene, resulting in substantial storage requirements. In contrast, our method not only enables fast optimization and rendering, but also achieves high fidelity with sparse views, while maintaining low storage requirements.

\subsection{Novel View Synthesis with Sparse-Views} 
While both NeRFs~\cite{mildenhall2021nerf} and 3DGS~\cite{kerbl20233d} achieve exceptional rendering quality with dense imagery, their performance degrades significantly with sparse views due to overfitting. This has spurred extensive research into improving their few-shot performance. 1) Sparse-View NeRF~\cite{yu2021pixelnerf,jain2021putting,niemeyer2022regnerf,wang2023sparsenerf,yang2023freenerf}. Before 3DGS, the NeRF community pioneered several solutions to this problem. These strategies include learning from external priors using pre-trained image encoders (pixelNeRF)~\cite{yu2021pixelnerf} or ~\cite{jain2021putting} designed a semantic consistency loss using CLIP embeddings ~\cite{radford2021learning}, enforcing geometric consistency through patch-based appearance and depth regularization (RegNeRF~\cite{niemeyer2022regnerf}, SparseNeRF~\cite{wang2023sparsenerf}), and applying frequency regularization to penalize high-frequency noise and improve generalization (FreeNeRF)~\cite{yang2023freenerf}. 2) Sparse-View 3DGS~\cite{chung2024depth,li2024dngaussian,paliwal2024coherentgs,zhang2024cor,zhu2024fsgs,park2025dropgaussian,xu2025dropoutgs}. More recently, a variety of innovative solutions have been developed for the 3DGS framework. Depth Regularization: One line of work focuses on resolving depth ambiguity using learnable parameters (Depth-GS)~\cite{chung2024depth} or local depth normalization (DN-Gaussian)~\cite{li2024dngaussian}. 
Consistency and Densification: Other methods improve multi-view consistency using optical flow (CoherentGS) ~\cite{paliwal2024coherentgs} or model disagreement (CoR-GS)~\cite{zhang2024cor}, while some adaptively densify sparse regions using "Gaussian unpooling" (FSGS)~\cite{zhu2024fsgs}.
Stochastic Regularization: Inspired by dropout, a third approach introduces stochasticity by randomly removing Gaussians during training to effectively mitigate overfitting (DropGaussian~\cite{park2025dropgaussian}, DropoutGS~\cite{xu2025dropoutgs}).
Different from the aforementioned methods, our approach addresses the sparse view issue at the representation level and proposes an LoD-structured 3DGS that enables robust reconstruction even with sparse views.

\subsection{Level-of-Detail (LoD)}  
In 3D rendering, LoD refers to the adjustment of the level of detail of 3D models based on their distance from the camera~\cite{nam2023mip,barron2023zip}. Several recent works ~\cite{cui2024letsgo,liu2024citygaussian,kerbl2024hierarchical,kulhanek2025lodge,ren2025octree} have explored LoD for 3D graphics systems (3DGS). LetsGo~\cite{cui2024letsgo} introduces LoD as a multi-resolution Gaussian representation for large-scale scene generation. CityGaussian~\cite{liu2024citygaussian} proposes a block-wise LoD, where during rendering, all Gaussians within the same block share the same level of detail, which is determined by the block’s distance from the camera.
Similarly, Hierarchical-GS~\cite{kerbl2024hierarchical} divides the large scene into distinct chunks and constructs a LoD for each chunk, facilitating hierarchy generation and consolidation for fast rendering. 
A more scalable approach was recently introduced by Octree-GS~\cite{ren2025octree}, which uses a unified octree with an accumulative LoD strategy to eliminate Gaussian redundancy.
Unlike these methods, we propose an anchor- and octree-based LoD-structured 3DGS representation with a hierarchical Gaussian drop technique to prevent overfitting. Additionally, our method specifically addresses the critical challenge of Gaussian storage in a streaming manner, which has not been considered by these approaches.

\section{Preliminatries}
\label{sec:Preliminatries}
\textbf{LoD-GS}~\cite{shen2025lod} achieves Level-of-Detail (LoD) rendering by organizing 3D Gaussians into hierarchical layers. Each layer $l$ contains Gaussians $\mathcal{G}_l$ with position $\boldsymbol{\mu}_l$, covariance matrix $\boldsymbol{\Sigma}_l$ (decomposed into scale $\boldsymbol{s}_l$ and quaternion $\boldsymbol{q}_l$), opacity $\boldsymbol{\alpha}_l$, and spherical harmonic coefficients $\boldsymbol{c}_l$, forming a multi-resolution structure: 
\begin{equation}\label{hierarchy}
\mathcal{G} = \{\mathcal{G}_0, \mathcal{G}_1, \ldots, \mathcal{G}_{L-1}\}, \quad |\mathcal{G}_l| < |\mathcal{G}_{l+1}|
\end{equation}
where coarser layers capture global structure, and finer layers encode details. During rendering, the active layer $l^*$ is selected based on viewing distance $d$:
\begin{equation}\label{lod_selection}
l^* = \min\left(\left\lfloor \log_\beta \frac{d}{d_0} \right\rceil ,L\right)
\end{equation}
with $d_0$ as the base distance and $\beta > 1$ controlling the transition rate. Only Gaussians in active layers are rasterized via tile-based rendering~\cite{kerbl20233d}, reducing computational cost for distant views while maintaining visual quality through distance-aware training and progressive densification.

\noindent\textbf{Scaffold-GS}~\cite{lu2024scaffold} 
introduces an anchor-based structured 3D Gaussian representation to reduce redundant Gaussians while maintaining high-quality rendering. 
The method begins by constructing a sparse voxel grid from Structure-from-Motion~\cite{snavely2006photo} (SfM)-derived points, with an anchor placed at the center of each voxel. Each anchor point is associated with a learnable feature $\hat{\mathbf{f}}$ and $K$ neighboring neural Gaussians, generated from the anchor using offsets:
\begin{equation}\label{nGaussians}
\{\boldsymbol{\mu}_0, \ldots, \boldsymbol{\mu}_{K-1}\} 
= \boldsymbol{\mu}_{\text{anchor}} + \{\mathbf{O}_0, \ldots, \mathbf{O}_{K-1}\}
\end{equation}
where $\boldsymbol{\mu}_{\text{anchor}}$ is the anchor center, $\mathbf{O}_i$ is a learnable offset. 
Other Gaussian attributes are directly decoded from the anchor feature $\hat{\mathbf{f}}$, the relative viewing distance $\Delta_{c}$ (the distance between the anchor center and the camera position), and the direction $\vec{d}_{c}$ (the vector from the camera to the anchor) through individual MLPs, denoted as $\text{F}_{\alpha}$, $\text{F}_{c}$, $\text{F}_{q}$, $\text{F}_{s}$. For instance, opacities are: 
\begin{equation}\label{mlps}
\{\boldsymbol{\alpha}_0, \ldots, \boldsymbol{\alpha}_{K-1}\} = \text{F}_{\alpha}(\hat{\mathbf{f}}, \Delta_{c}, \vec{d}_{c})
\end{equation}

\begin{figure*}[h]
\centering
\includegraphics[width=0.999\textwidth]{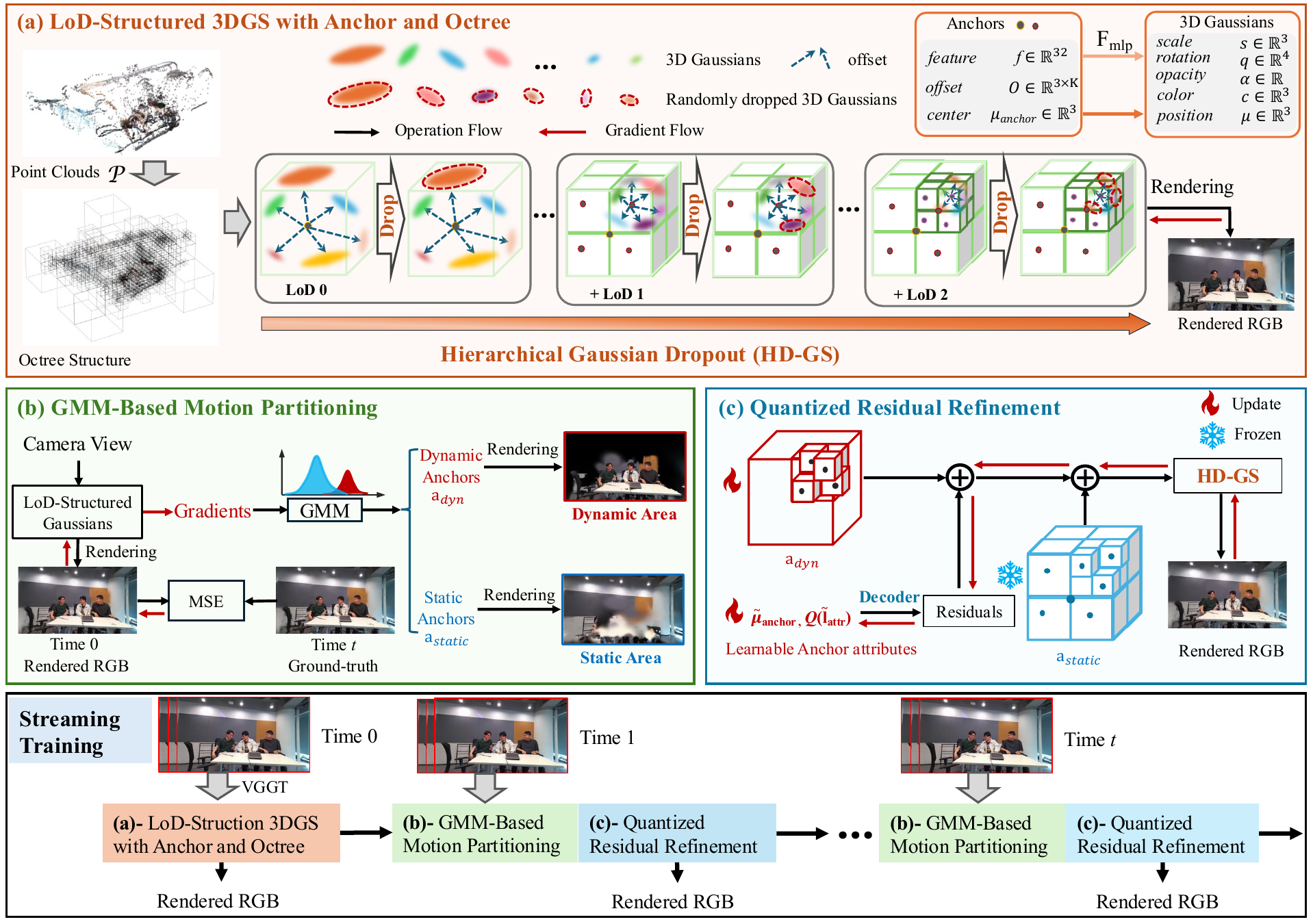} 
\caption{Illustration of our proposed StreamLoD-GS. It comprises three major components: \textbf{(a)} LoD-Structured 3DGS with Anchor and Octree; \textbf{(b)} GMM-Based Motion Partitioning; \textbf{(c)} Quantized Residual Refinement. \textbf{Bottom:} The streaming training pipeline. Frames are processed sequentially: the initial frame (Time 0) undergoes LoD-structured 3DGS optimization, while subsequent frames apply motion partitioning and residual refinement to ensure temporally coherent RGB rendering throughout the stream. }
\label{fig:method}
\end{figure*}

\section{Methodology}
\label{sec:method}
Given a sequence of videos captured by $N$ synchronized cameras, denoted as $\{\mathbf{I}^{(i)}_t \}_{i=1}^{N}$ for $t = 0, 1, \dots, T$, our goal is to reconstruct the video scene in a streaming manner.
To achieve this, we propose StreamLoD-GS, a hierarchical LoD-structured Gaussian framework, as shown in Fig.~\ref{fig:method}.
We begin by initializing an anchor- and octree-based LoD-structured 3DGS representation using the multi-view images from the first frame $\{ \mathbf{I}^{(i)}_{t=0} \}_{i=1}^{N}$  (Sec.~\ref{sec:s1}); Next, we design a GMM-based motion partitioning method to identify dynamic anchors for selective modeling, while freezing static regions to ensure background stability and reduce training costs (Sec.~\ref{sec:s2}); Finally, we propose a quantized residual refinement method to quantize the residuals of the dynamic anchors to reduce storage costs (Sec.~\ref{sec:s3}).
 
\subsection{LoD-Structured 3DGS with Anchor and Octree}
\label{sec:s1} 
Upon receiving the multi-view images at the first frame $\{ \mathbf{I}^{(i)}_{t=0} \}_{i=1}^{N}$, we first use them to initialize our LoD-structured representation. We then apply a hierarchical Gaussian dropout to progressively drop Gaussian at each layer to prevent overfitting.

\textbf{Anchor and LoD Initialization.} 
Standard point cloud initialization typically employs COLMAP-based SfM~\cite{schonberger2016structure}; however, its reliability diminishes under Sparse-View  constraints. To circumvent this, we utilize VGGT~\cite{wang2025vggt} to estimate the initial point cloud $\mathcal{P}$ from the first frame's multi-view images $\{\mathbf{I}^{(i)}_{t=0} \}_{i=1}^{N}$, providing a more accurate starting point for subsequent optimization. Crucially, to ensure a fair experimental comparison, we apply this consistent initialization across all evaluated methods in Sec.~\ref{sec:experiments}. Following the acquisition of $\mathcal{P}$, we initialize the Gaussian centers and organize them into an octree structure with an $L$-layered LoD hierarchy, following Octree-GS~\cite{ren2024octree}. 
Here $L$ is determined by the scene scale bounds $(d_{\max}, d_{\min})$ as $L = \lfloor \log_2(d_{\max}/d_{\min}) \rceil + 1$, where $\lfloor.\rceil$ denotes the round operator.
Instead of directly placing an anchor in each voxel, we perform a layer-wise placement. 
At each level $l \in \{0, ..., L-1\}$, an anchor is placed in a voxel of size $\delta_l = \delta / 2^l$, where $\delta$ is the base voxel size at the coarsest level (level 0). Then similar to Scaffold-GS~\cite{lu2024scaffold}, each anchor is assigned a feature $\hat{\mathbf{f}}$ and is associated with $K$ neural Gaussian, each having a learnable offset for Gaussian positions.

To further reduce computational costs, we employ a dynamic anchor selection strategy~\cite{ren2025octree} to select the most important anchors during rendering. 
For an arbitrary anchor $\mathbf{a}_j$, its original layer index can be computed as $L_j=\lfloor\log_2 \left( \frac{d_{\text{max}}}{\Delta_{c_j}} \right)\rceil$, where $\Delta c_j$ is the distance between the anchor's center and the camera position.
During training, its gradient regarding to the 2D screen-space projection position is recorded as $\mathbf{g}_{\boldsymbol{\mu}_{\text{j}}}$. Then its new layer index can be updated by:
\begin{equation}
L_{j}^* = \lfloor\log_2 \left( \frac{d_{\text{max}}}{\Delta_{c_j}} \right)\rceil + \Delta L, if ~\mathbf{g}_{\boldsymbol{\mu}_{\text{j}}} > \mathbf{g}_{\text{threshold}}
\end{equation}
where $\mathbf{g}_{\text{threshold}}$ is a threshold.
During progressive training, we limit the maximum level to $L_{\max}$. An anchor is selected for rendering if:
\begin{equation}
L_j \leq \min(\lfloor L_{j}^* \rceil, L_{\max})
\end{equation}
This function implies that if an anchor moves to a distant coarser layer, it is considered unimportant. As a result, we remove it from the rendering process.

\noindent\textbf{Hierarchical Gaussian Dropout.} 
To mitigate overfitting caused by limited training viewpoints and stabilize the optimization process, we enhance DropGaussian~\cite{park2025dropgaussian} by introducing a level-aware dropout strategy.
Since higher LoDs comprise denser Gaussians, we apply a progressively increasing dropout rate defined as $r^{(l)}_m = \gamma^{(l)} \cdot \frac{m}{M}$, where $m$ is the current training step, $M$ is the total number of training steps, and the scale factor $\gamma^{(l)} = 0.1 + 0.05 \cdot l$ grows with the LoD level $l$.

Finally, unlike Scaffold-GS~\cite{lu2024scaffold}  that uses separate networks to predict Gaussian attributes, we use a shared MLP network $\text{F}_{\text{mlp}}$ to predict Gaussian attributes of the hierarchical Gaussians $\mathcal{G} = \{\mathcal{G}_0, \mathcal{G}_1, \ldots, \mathcal{G}_{L-1}\}$ from these anchor features $\hat{\mathbf{f}}$, viewing distance $\Delta_{c}$, and camera information $\vec{d}_{c}$: 
\begin{equation}\label{uni_mlp}
\mathcal{G} = \{\mathcal{G}_0, \mathcal{G}_1, \ldots, \mathcal{G}_{L-1}\} = \text{F}_{\text{mlp}}(\hat{\mathbf{f}}, \Delta_{c}, \vec{d}_{c})
\end{equation}
We find that using a shared MLP helps avoid the introduction of excessive parameters from multiple networks, thereby accelerating both training and rendering without degrading performance.

\subsection{GMM-based Motion Partitioning}
\label{sec:s2}
After initializing the LoD-Structured Gaussian with the first frame (defined as the canonical space $\mathcal{G}_{\text{cano}}$) as described in Sec.~\ref{sec:s1}, we focus on modeling only the dynamic regions in subsequent frames to reduce optimization costs and enable faster training. 
Our idea is to compare the current frame $\{ \mathbf{I}^{(i)}_{t=c} \}_{i=1}^{N}$ with the canonical frame $\{ \mathbf{\hat{I}}^{(i)}_{t=0} \}_{i=1}^{N}$ rendered from $\mathcal{G}_{\text{cano}}$ to calculate the gradient of their differences to estimate the spatial variations. This gradient will allow us to categorize anchors of the scene into two groups: dynamic anchors $\mathbf{a}_{\text{dyn}}$ and static anchors $\mathbf{a}_{\text{static}}$. The gradient is calculated by:
\begin{equation}
\mathbf{g}_{\boldsymbol{\mu}_{\text{cano}}}=\frac{1}{N} \sum_{i=1}^{N}
\nabla_{\boldsymbol{\mu}_{\text{cano}}} \left\| \mathbf{I}^{(i)}_{t=c} \ - \mathbf{\hat{I}}^{(i)}_{t=0} \right\|^2
\end{equation}
where $\boldsymbol{\mu}_{\text{cano}}$  denotes the 2D screen-space projection of the Gaussian center $\boldsymbol{\mu}$ in the canonical space.
We then apply a GMM~\cite{permuter2006study} to cluster anchors based on the gradient $\mathbf{g}_{\boldsymbol{\mu}_{\text{cano}}}$, where the component with the higher mean corresponds to the dynamic regions.
Anchors are classified as dynamic anchors $\mathbf{a}_{\text{dyn}}$ if their GMM posterior probability exceeds the threshold $\rho = 80\%$, while the remaining anchors are classified as static anchors $\mathbf{a}_{\text{static}}$.
The dynamic anchors $\mathbf{a}_{\text{dyn}}$ are subsequently utilized for quantized dynamics learning, as described in Sec.~\ref{sec:s3}.

\subsection{Quantized Residual Refinement}
\label{sec:s3}
After decomposing the dynamic anchors $\mathbf{a}_{\text{dyn}}$ and static ones  $\mathbf{a}_{\text{static}}$, we aim to efficiently compress the temporal changes in dynamic scenes. This is accomplished by learning a quantized residual $\tilde{\mathbf{r}}$ for the dynamic anchors $\mathbf{a}_{\text{dyn}}$: 
\begin{equation}
\mathbf{A}_{\text{dyn}} = \mathbf{A}_{\text{dyn}} + \tilde{\mathbf{r}}, \quad \tilde{\mathbf{r}} = 
\begin{cases}
\tilde{\boldsymbol{\mu}}_{\text{anchor}} & \text{Anchor location} \\
\mathbf{Q}(\tilde{\mathbf{l}}_{\text{attr}}) & \text{Other attributes}
\end{cases}
\end{equation}
where $\mathbf{A}_{\text{dyn}}$ represents the Anchor attributes of $\mathbf{a}_{\text{dyn}}$. 
$\tilde{\boldsymbol{\mu}}_{\text{anchor}}$ is a learnable parameter representing the Anchors' spatial coordinates, which we do not quantize due to its negligible storage footprint. 
$\tilde{\mathbf{l}}_{\text{attr}}$ is also a learnable parameter (comprising the anchor feature and offset) that we subject to quantization. We define $\mathbf{Q}(\tilde{\mathbf{l}}_{\text{attr}})$ as the quantized latents, with the quantization optimized through the Straight-Through Estimator \cite{bengio2013estimating}. 
By fixing the static anchors $\mathbf{a}_{\text{static}}$ and modeling the quantized residuals of the dynamic anchors $\mathbf{a}_{\text{dyn}}$, our method achieves approximately 80\% storage reduction while preserving visual quality (see Tab.~\ref{table:ablation}).

\subsection{Streaming Training} 
\label{sec:s4}
Fig.\ref{fig:method}-bottom illustrates our streaming training process. In the first step, we initialize the LoD-structured 3D Gaussian representation (Fig.\ref{fig:method}-(a)) using the first frame $\{ \mathbf{I}^{(i)}_{t=0} \}_{i=1}^{N}$ to train the model. This step determines the structure of the canonical space $\mathcal{G}_{\text{cano}}$, including the number of LoD layers, the initial Gaussian attributes, and the centers and quantity of anchors.
For subsequent frames $\{\mathbf{I}^{(i)}_t \}_{i=1}^{N}$ where $t = 1, 2, \dots, T$, we optimize both GMM-based Motion Partitioning and Quantized Residual Refinement in a frame-by-frame, streaming manner, where only the parameters related to dynamic anchors are optimized. All training is conducted using the same loss function:
\begin{equation}
\mathcal{L}_{\text{recon}} = (1 - \lambda) \mathcal{L}_1 + \lambda \mathcal{L}_{D\text{-SSIM}}
\end{equation}
where $\lambda$ is a hyper-parameter. We set $\lambda = 0.2$ to prioritize sharp reconstructions while maintaining perceptual coherence~\cite{sun20243dgstream}.

\section{Experiments}
\label{sec:experiments}
In this section, we first introduce the experimental setup. Next, we present quantitative and qualitative comparisons against state-of-the-art benchmarks to validate our reconstruction quality across various view densities. Furthermore, we provide an systematic ablation analysis targeting individual components, the number of training views, and the LoD-AO farchitecture—concluding with an evaluation of the limitations of the proposed StreamLoD-GS.
 
\subsection{Experimental Setup}  
\subsubsection{Datasets}
We conduct experiments on two public datasets: the Neural 3D Video (\textbf{N3DV})~\cite{li2022neural} and the \textbf{Meet Room}~\cite{li2022streaming} datasets. N3DV comprises six dynamic indoor scenes, where each video is captured at a resolution of $2704 \times 2078$ and 30 FPS, with 18-21 videos per scene. Following~\cite{sun20243dgstream,gao2024hicom,yan2025instant,girish2024queen,hu20254dgc}, we downsample the resolution to $1352 \times 1014$. We adopt 3, 4, and 6 views for training and use the remaining views for evaluation.

The {Meet Room} dataset was recorded using 13 synchronized Azure Kinect cameras across 3 indoor scenes, namely ``Discussion'', ``Trimming'', and ``Vrheadset'', at a resolution of $1280 \times 720$ and 30 FPS. We use the original resolution and train with 2, 3, 4, 5, 6, 9, and 12 views, using remaining views for evaluation.

\subsubsection{Implementation} 
Following the protocols in ~\cite{sun20243dgstream,gao2024hicom,yan2025instant,girish2024queen,hu20254dgc}, we utilize the first 300 frames (around 10 seconds) for evaluation. 
Initial point clouds are generated using VGGT~\cite{wang2025vggt} based on the provided camera poses. To ensure a rigorous and fair comparison, this initialization is consistently applied across our framework and all competing baselines for every scene. 
Regarding model hyper-parameters, we set the base voxel size $\delta$ to 0.001 and and the number of Gaussians per anchor $K$ to 10. Based on empirical observations, the GMM posterior probability threshold $\rho$ is established at $80\%$. The $F_\text{mlp}$ module (detailed in Fig.~\ref{fig:method} and Sec.~\ref{sec:s1}) comprises 2-layer MLPs with ReLU activation and a 32-dimensional hidden layer. For attribute quantization, the latent dimensionality for both anchor features and offsets is uniformly set to 12. As specified in Sec.~\ref{sec:s1}, the maximum LoD level
$L_{\max}$ is incrementally increased during optimization. 
For anchor point refinement, we compute average gradients over 200 iterations for the initial frame and 30 iterations for subsequent frames. Following each refinement cycle, an anchor is pruned if the accumulated opacity of its associated neural Gaussians falls below a threshold of 0.05. All remaining hyper-parameters adhere to the configurations specified in Octree-GS~\cite{ren2025octree}. 
 
The model is trained for 500 epochs during the initial timestep, followed by 10 epochs for each subsequent timestep to facilitate rapid adaptation.  Following the evaluation protocols established in~\cite{sun20243dgstream,gao2024hicom,yan2025instant,girish2024queen}, we assess visual quality using frame-wise Peak Signal-to-Noise Ratio (PSNR), Structural Similarity Index (SSIM)~\cite{wang2004ssim}, and Learned Perceptual Image Patch Similarity (LPIPS)~\cite{zhang2018LPIPS}. These metrics are averaged across all views to ensure a comprehensive evaluation. Additionally, we report the average storage footprint, training latency per timestep, and rendering throughput (FPS). All experiments are conducted on a single NVIDIA A800-SXM4-80GB GPU.

\begin{table*}[!htbp]
\centering
\caption{Quantitative Comparison in Meet Room~\cite{li2022streaming} dataset with 3, 4, and 5 training views. The best results within each category is marked in \textbf{bold}. The second results are highlighted with an \underline{underline}. }
\label{table:meetingroom}
\renewcommand{\arraystretch}{1.07}
\begin{tabular}{llcccccc}
    \hline
    \textbf{Views} & \textbf{Methods} & PSNR (dB)$\uparrow$ & SSIM$\uparrow$ & LPIPS$\downarrow$ & Storage (MB)$\downarrow$ & Train (s)$\downarrow$ & Render (FPS)$\uparrow$ \\
    \hline
    \multirow{7}{*}{3-views} 
    & StreamRF~\cite{li2022streaming} & 15.82 & 0.641 & 0.510 & 22.81 & 7.854 & 14.9 \\
    & 3DGStream~\cite{sun20243dgstream} & 17.24 & 0.687 & 0.481 & 3.822 & 5.526 & \underline{588} \\
    & HiCoM~\cite{gao2024hicom} & 16.92 & 0.665 & 0.431 & 0.913 & 1.762 & 517 \\
    & 4DGC~\cite{hu20254dgc} & 17.36 & 0.713 & 0.416 & 0.653 & 17.41 & 418 \\
    & QUEEN~\cite{girish2024queen} & 17.01 & 0.670 & 0.443 & 0.860 & \textbf{0.455} & 486 \\
    & StreamLoD-GS$\star$ & \underline{18.50} & \underline{0.783} & \underline{0.400} & \textbf{0.245} & \underline{0.643} & \textbf{677} \\
    & StreamLoD-GS & \textbf{18.85} & \textbf{0.793} & \textbf{0.383} & \underline{0.253} & 1.073 & 547 \\
    \hline
    \multirow{7}{*}{4-views} 
    & StreamRF~\cite{li2022streaming} & 16.75 & 0.748 & 0.452 & 19.53 & 8.210 & 17.1 \\
    & 3DGStream~\cite{sun20243dgstream} & 18.36 & 0.797 & 0.413 & 3.850 & 5.510 & 609 \\
    & HiCoM~\cite{gao2024hicom} & 18.58 & 0.809 & 0.391 & 0.894 & 2.012 & 524 \\
    & 4DGC~\cite{hu20254dgc} & 19.45 & 0.810 & 0.385 & 0.794 & 18.61 & 441 \\
    & QUEEN~\cite{girish2024queen} & 19.26 & 0.753 & 0.386 & 0.823 & \textbf{0.581} & 484 \\
    & StreamLoD-GS$\star$ & \underline{20.67} & \underline{0.823} & \underline{0.370} & \textbf{0.209} & \underline{0.823} & \textbf{689} \\
    & StreamLoD-GS & \textbf{21.40} & \textbf{0.832} & \textbf{0.351} & \underline{0.215} & 1.581 & \underline{625} \\
    \hline
    \multirow{7}{*}{5-views} 
    & StreamRF~\cite{li2022streaming} & 17.57 & 0.780 & 0.432 & 16.46 & 8.921 & 19.3 \\
    & 3DGStream~\cite{sun20243dgstream} & 19.53 & 0.817 & 0.400 & 3.841 & 5.440 & \underline{619} \\
    & HiCoM~\cite{gao2024hicom} & 19.69 & 0.826 & 0.391 & 0.819 & 2.225 & 518 \\
    & 4DGC~\cite{hu20254dgc} & 21.28 & 0.831 & 0.380 & 0.639 & 19.52 & 425 \\
    & QUEEN~\cite{girish2024queen} & 21.22 & 0.803 & \underline{0.343} & 0.603 & \textbf{0.692} & 485 \\
    & StreamLoD-GS$\star$ & \underline{21.95} & \underline{0.843} & \underline{0.343} & \textbf{0.198} & \underline{0.973} & \textbf{621} \\
    & StreamLoD-GS & \textbf{22.73} & \textbf{0.861} & \textbf{0.310} & \underline{0.205} & 1.910 & 608 \\
    \hline
\end{tabular}
\end{table*}

\begin{table*}[!htbp]
\centering
\caption{Quantitative Comparison in N3DV~\cite{li2022neural} dataset with 3, 4, and 5 training views. The best results within each category is marked in \textbf{bold}. The second results are highlighted with an \underline{underline}. }
\renewcommand{\arraystretch}{1.07}
\begin{tabular} {llcccccc } 
    \hline
    \textbf{Views} &\textbf{Methods} & PSNR (dB)$\uparrow$ & SSIM$\uparrow$ & LPIPS$\downarrow$ & Storage (MB)$\downarrow$ & Train (s)$\downarrow$ & Render (FPS)$\uparrow$ \\
    \hline
    \multirow{7}{*}{3-views} 
    &StreamRF~\cite{li2022streaming} &18.79 & 0.732 & 0.401 &41.23 &13.614  & 19.5  \\
    &3DGStream~\cite{sun20243dgstream} & 19.89& 0.747 & 0.373 & 6.301 & 9.162 & \textbf{662}\\
    &HiCoM~\cite{gao2024hicom} & 19.21& 0.741 & 0.352 &1.201 & 1.83 & 512 \\
    &4DGC~\cite{hu20254dgc} &19.82 & 0.763&0.339 & 0.871&17.49 & 484\\
    &QUEEN~\cite{girish2024queen} &19.25 & 0.743& 0.330 &0.962 & \textbf{0.452} & 460 \\
    &StreamLoD-GS$\star$ &\underline{20.50} & \underline{0.785} & \underline{0.320} & \textbf{0.112} & \underline{0.793}  &\underline{651}  \\
    &StreamLoD-GS & \textbf{20.54}& \textbf{0.790} & \textbf{0.316} & \underline{0.126}& 1.336& 530 \\
    \hline
    \multirow{7}{*}{4-views}
    &StreamRF~\cite{li2022streaming} & 19.08& 0.747 & 0.390 & 39.68& 16.46 &  22.3 \\
    &3DGStream~\cite{sun20243dgstream} & 20.16& 0.780 & 0.363 & 6.610 & 8.830 & \textbf{687} \\
    &HiCoM~\cite{gao2024hicom} & 19.81 & 0.767 & 0.326 & 1.024& 2.36 & 501  \\
    &4DGC~\cite{hu20254dgc} &20.12 & 0.779&0.305 &0.837 & 18.20& 478\\
    &QUEEN~\cite{girish2024queen}&19.64 & 0.773 & 0.293 & 0.710& \textbf{0.568} & 482 \\
    &StreamLoD-GS$\star$ & \underline{21.14}& \underline{0.812} & \underline{0.283} & \textbf{0.243} & \underline{0.982} & \underline{679} \\
    &StreamLoD-GS & \textbf{21.30} & \textbf{0.817} & \textbf{0.280} & \underline{0.256} & 1.696 & 525 \\
    \hline
    \multirow{7}{*}{6-views} 
    &StreamRF~\cite{li2022streaming} &20.95 & 0.762 & 0.380 & 12.78& 17.25 &  26.1  \\
    &3DGStream~\cite{sun20243dgstream} &21.85 & 0.800 & 0.351 & 6.624& 8.965 & \textbf{666} \\
    &HiCoM~\cite{gao2024hicom} & 21.17 & 0.812 & 0.306 &0.847 & 4.519 &  499 \\
    &4DGC~\cite{hu20254dgc} & 22.54& 0.833& 0.298& 0.784& 19.07& 472 \\
    &QUEEN~\cite{girish2024queen}&22.88 & 0.841 & 0.289 &0.718 & \textbf{0.816} & 466 \\
    &StreamLoD-GS$\star$ & \underline{23.36}& \underline{0.842} & \underline{0.271} & \textbf{0.211}& \underline{1.401} & \underline{524}  \\
    &StreamLoD-GS &\textbf{23.85} &  \textbf{0.852}& \textbf{0.256} &\underline{0.223} & 2.313 & 506 \\
    \hline
\end{tabular} 
\label{table:dynerf}
\end{table*}

\begin{figure*}[h]
\centering
\includegraphics[width=0.996\textwidth]{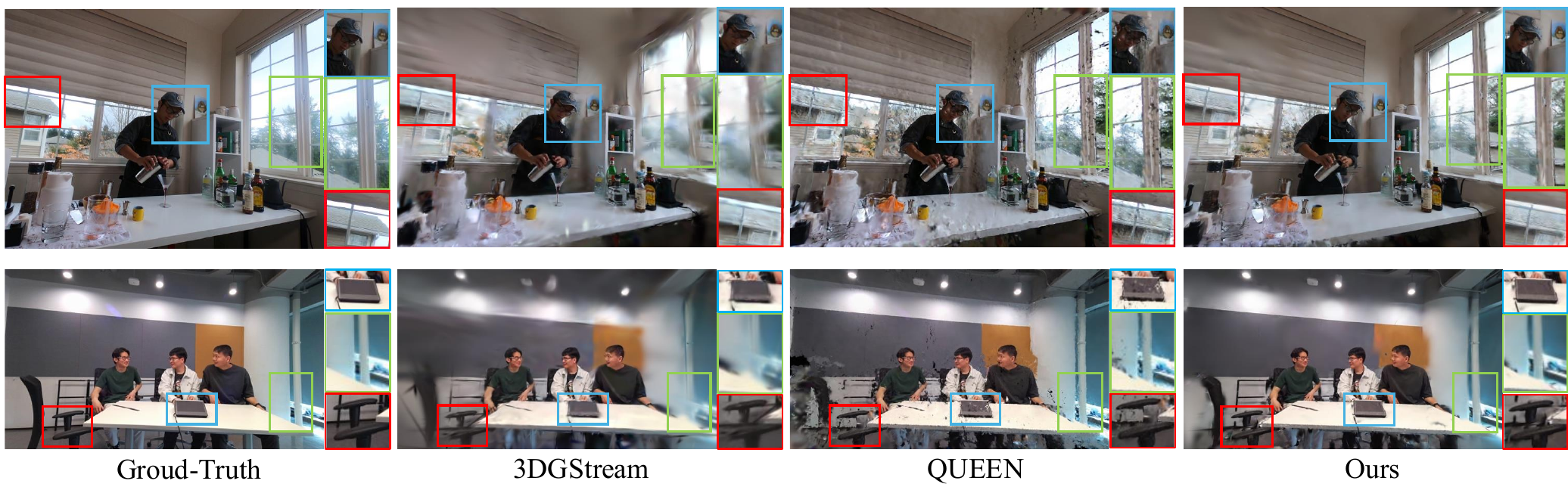} 
\caption{Qualitative Results in N3DV~\cite{li2022neural} and Meet Room~\cite{li2022streaming} Datasets with 3 training views. We demonstrate novel view results produced by 3DGStream~\cite{sun20243dgstream}, QUEEN~\cite{girish2024queen}, and our approach for comparison. }
\label{fig:Visualization}
\end{figure*}

\subsubsection{Baseline Methods}
We evaluate the proposed StreamLoD-GS against several state-of-the-art SFVV frameworks, including StreamRF~\cite{li2022streaming}, 3DGStream~\cite{sun20243dgstream}, HiCoM~\cite{gao2024hicom}, 4DGC~\cite{hu20254dgc}, and QUEEN~\cite{girish2024queen}. 

To ensure a rigorous and unbiased evaluation, all baselines are implemented using their official configurations within a standardized experimental environment. Specifically, we enforce a consistent point cloud initialization (utilizing VGGT~\cite{wang2025vggt}) across all methods for each scene.
For QUEEN, which adopts a similar training protocol to ours, adhere strictly to its official settings.
All models are retrained using identical view inputs on the same hardware to eliminate any performance discrepancies arising from computational variances.
Additionally, we evaluate a specialized variant, StreamLoD-GS$^\star$, derived from our proposed StreamLoD-GS. This variant performs static anchor updates and dynamic anchor identification (as described in Sec.~\ref{sec:s2}) every four frames to accumulate richer gradient information. Moreover, given the limited number of dynamic anchors within short intervals, StreamLoD-GS$^\star$ classifies an anchor as dynamic when its posterior probability, as defined in Sec.~\ref{sec:s2}, exceeds $\rho = 90\%$.

\subsection{Quantitative Comparisons under Sparse Views}
We conduct a comprehensive quantitative evaluation across varying view densities, comparing our framework against state-of-the-art baselines in Tab.~\ref{table:meetingroom} and Tab.~\ref{table:dynerf}.
Tab.~\ref{table:meetingroom} details the comparative results on the Meeting Room~\cite{li2022streaming} dataset. The results demonstrate that StreamLoD-GS consistently surpasses nearly all baselines across the tested view configurations. Specifically, in the highly challenging 3-view setting, StreamLoD-GS achieves the highest quality metrics, while the StreamLoD-GS$\star$ variant maintains superior rendering speeds, the most compact storage, and competitive training efficiency. At 4 views, StreamLoD-GS reaches a PSNR of 21.40 dB, marking a significant 1.95 dB improvement over 4DGC while requiring only 27\% of its storage footprint. For the 5-view configuration, our method again secures the best quality metrics with a mere 0.205 MB of storage—66\% reduction compared to QUEEN, despite the latter's specialized Gaussian-based learned quantization framework.

Similarly, quantitative results on the N3DV dataset~\cite{li2022neural} (Tab.~\ref{table:dynerf}) further validate our model's efficiency.
With 3 training views, our method achieves a PSNR of 20.54 dB, outperforming the second-best method (4DGC) while consuming only 15\% of its storage. At 4 views, StreamLoD-GS reaches 21.30 dB, surpassing the nearest competitor by 1.14 dB. Notably, our storage footprint remains exceptionally lean at 0.256 MB, which is 64\% smaller than that of QUEEN. In the 6-view scenario, our approach maintains its lead in quality metrics with a compact 0.223 MB storage. Throughout these tests, the StreamLoD-GS$\star$ variant consistently ranks second in quality while delivering near-optimal rendering speeds and significantly accelerated training times.

\begin{table}[!htbp]
\centering
\caption{Quantitative Comparison in Meet Room~\cite{li2022streaming} dataset with 12 views for training and reserved one for test. The best results within each category is marked in \textbf{bold}. The second results are highlighted with an \underline{underline}. }
\begin{tabular}{lcccc}
    \hline
    \multirow{2}{*}{\textbf{Methods}} & PSNR & Storage & Train & Render \\
    & (dB)$\uparrow$ & (MB)$\downarrow$ & (s)$\downarrow$ & (FPS)$\uparrow$ \\
    \hline
    3DGStream~\cite{sun20243dgstream} & 26.36 & 4.51 & 4.95 & \underline{309}  \\
    HiCoM~\cite{gao2024hicom} &26.73 & 0.60 & 3.92 & 289  \\
    4DGC~\cite{hu20254dgc}& 26.87 & 0.63 & 21.3  &  273 \\
    QUEEN~\cite{girish2024queen} &\underline{27.35} &  \underline{0.35} & \textbf{1.31} & 254   \\
    StreamLoD-GS & \textbf{27.84}& \textbf{0.19} & \underline{2.93} &  \textbf{312} \\
    \hline
\end{tabular} 
\label{table:denseview}
\end{table}

\begin{table*}[!htbp]
\centering
\caption{Ablation Study Results on Meet Room~\cite{li2022streaming} dataset with 3 training views and reserved all for test. LoD-AO: LoD-Structured 3DGS with Anchor and Octree. GMM-Part: GMM-Based Motion Partitioning. Q: Quantized Residual Refinement; The best results within each category is marked in \textbf{bold}. }
\renewcommand{\arraystretch}{1.07}
\begin{tabular} {c|c|c|c|c|c|c|c|c }
    \hline
    LoD-AO & GMM-Part& Q &  PSNR (dB)$\uparrow$ & SSIM$\uparrow$ & LPIPS$\downarrow$ &  Storage (MB)$\downarrow$ & Train (s)$\downarrow$ & Render (FPS)$\uparrow$\\
    \hline
      \checkmark  & \checkmark & &  22.51&  0.852& 0.320 & 1.683 & 1.673 & 582\\
      \checkmark  &  &\checkmark   & \textbf{23.11} & 0.857 & 0.326 & 0.283& 2.197 & 425\\
        & \checkmark & \checkmark &  20.02& 0.778 & 0.337 & 0.633& \textbf{0.902} & 487 \\
      \checkmark  & \checkmark &\checkmark  & 22.73& \textbf{0.865} &  \textbf{0.310} & \textbf{0.205}& 1.910 & \textbf{608} \\
    \hline
\end{tabular} 
\label{table:ablation}
\end{table*}

\begin{figure*}[!htbp]
\centering
\includegraphics[width=0.99\textwidth]{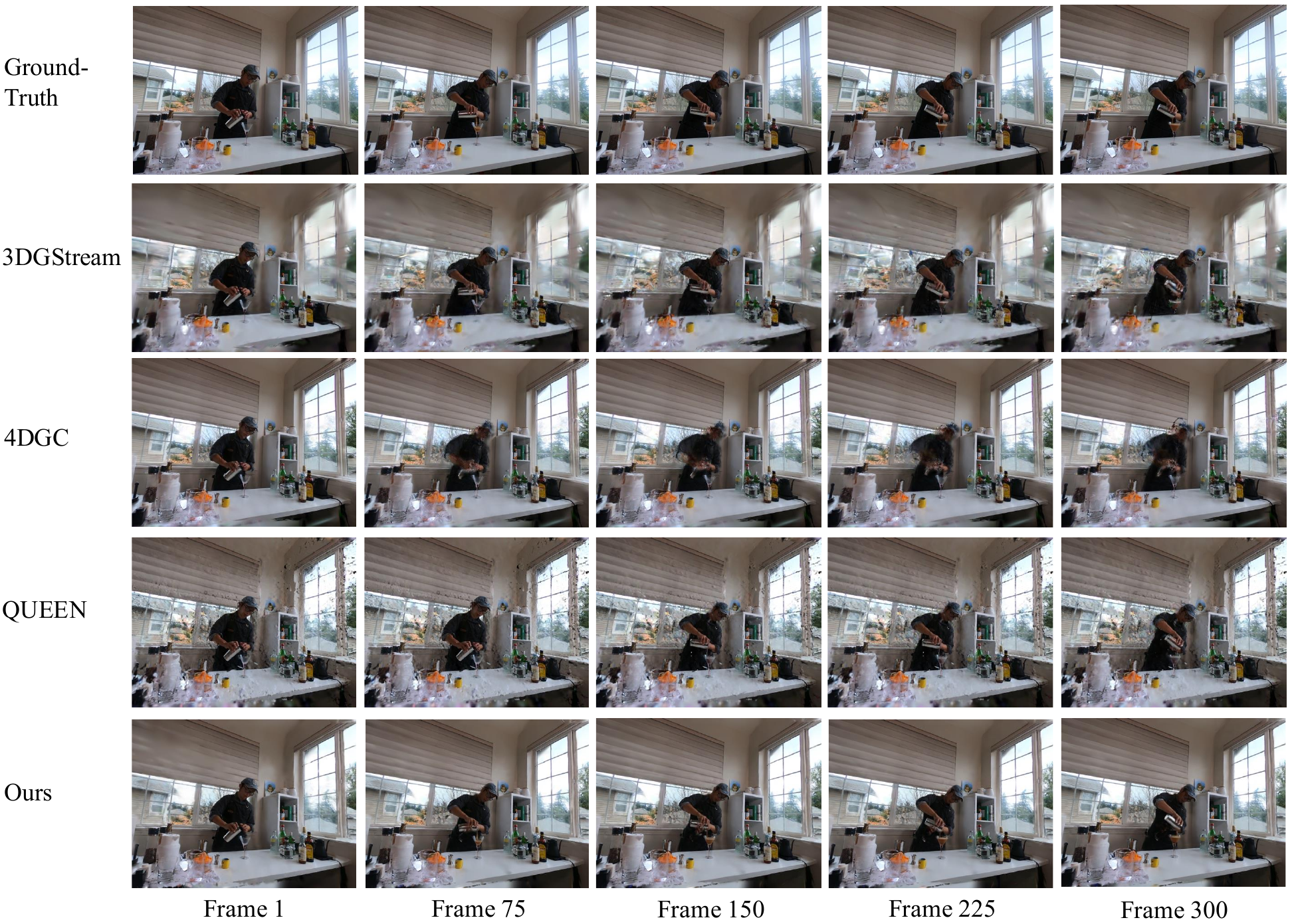} 
\caption{Quantitative comparison on the N3DV (Scene: Coffee Martini) dataset~\cite{li2022neural} with 3 training views. We compare novel view results on several frames sampled at equal temporal intervals, which are produced by 3DGStream~\cite{sun20243dgstream}, 4DGC~\cite{hu20254dgc}, QUEEN~\cite{girish2024queen}, and our approach.}
\label{fig:framelevel_N3DV}
\end{figure*}  

\subsection{Quantitative Comparisons under Dense Views}
To further evaluate the robustness of StreamLoD-GS, we conduct a comparative analysis against state-of-the-art baselines under dense-view conditions, utilizing 12 views for training and one for testing. 
As summarized in Tab.~\ref{table:denseview}, StreamLoD-GS consistently outperforms all competitors, achieving the highest PSNR, the fastest rendering throughput, and the most compact storage footprint, while maintaining highly competitive training efficiency. These results underscore the superior scalability and computational efficiency of our framework, demonstrating that StreamLoD-GS not only excels in Sparse-View reconstruction but also maintains a significant performance edge in information-rich, dense-view scenarios.

\subsection{Qualitative Analysis}
Fig.~\ref{fig:Visualization} presents a visual comparison of the reconstruction results across various methods.
On the N3DV dataset~\cite{li2022neural} (Scene: Coffee Martini, first row), our LoD-based method reconstructs finer and more structurally complete details than 3DGStream~\cite{sun20243dgstream} and QUEEN~\cite{girish2024queen}, \emph{e.g.}, the window and the transition between the person's head and the background wall. Notable improvements are visible in the crispness of the window frames and the sharp boundary transition between the subject's head and the background wall. Specifically, our approach effectively mitigates the pronounced floaters prevalent in 3DGStream and the structural erosion seen in QUEEN.
Similarly, for the Meeting Room dataset~\cite{li2022streaming} (Scene: Discussion, second row), our method preserves superior detail in both the foreground (\emph{e.g.}, the intricate structures of the chair and microphone) and the background (\emph{e.g.}, the reflections on the glass wall), whereas baseline methods exhibit significant blurring or geometric distortion. 
 
\begin{figure*}[!htbp]
\centering
\includegraphics[width=0.996\textwidth]{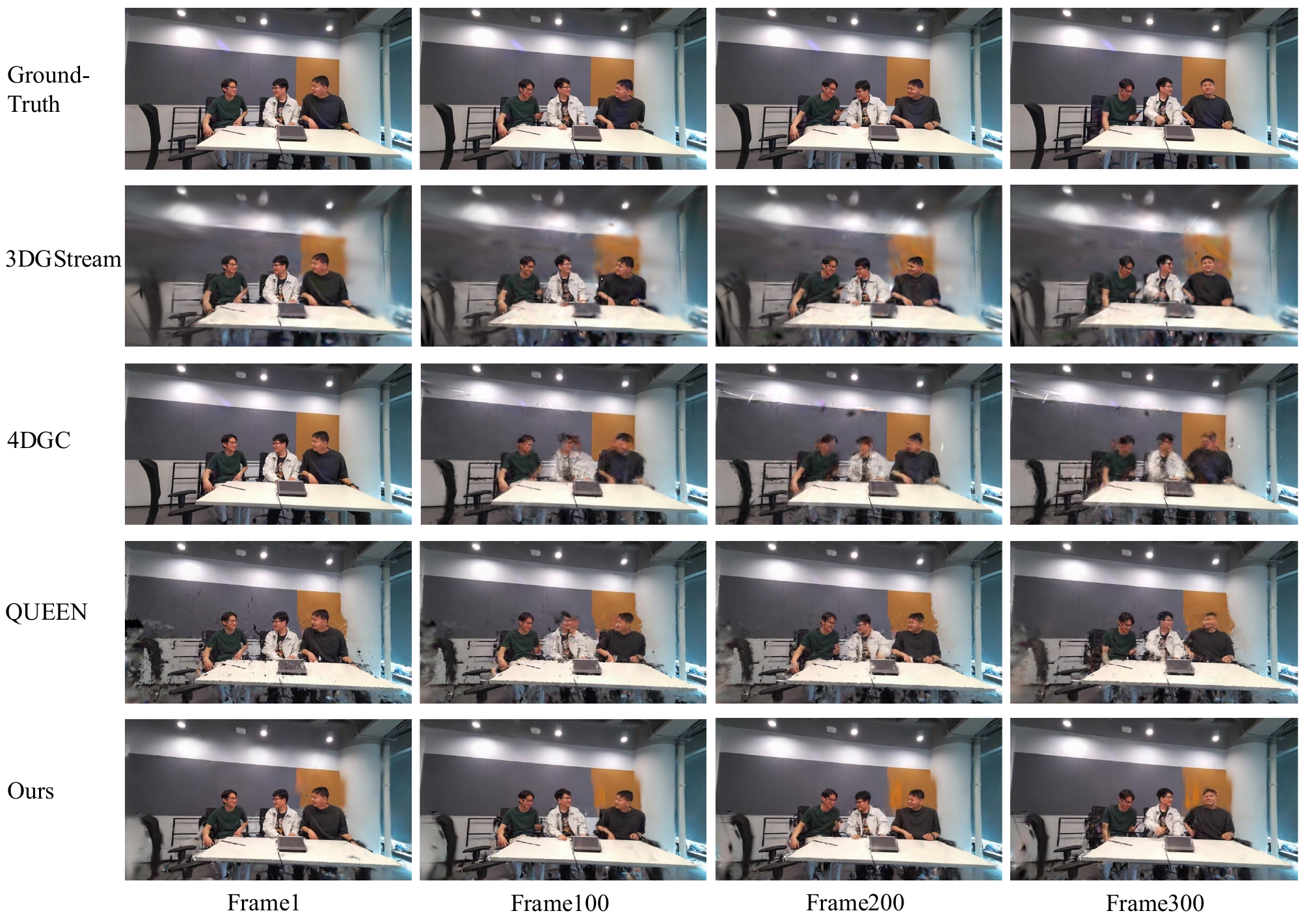} 
\caption{Quantitative comparison on the Meet Room (Scene: Discussion) dataset~\cite{li2022streaming} with 3 training views. We compare novel view results on several frames sampled at equal temporal intervals, which are produced by 3DGStream~\cite{sun20243dgstream}, 4DGC~\cite{hu20254dgc}, QUEEN~\cite{girish2024queen}, and our approach.}
\label{fig:framelevel_meetroom}
\end{figure*}

Additionally, we also compare the image details of the same frame across the current state-of-the-art  methods in the "Discussion" and "Coffee Martini" scenes. To provide a more comprehensive comparison, in Fig.\ref{fig:framelevel_N3DV} and Fig.\ref{fig:framelevel_meetroom}, we compare the reconstruction results of different methods across several frames sampled at equal temporal intervals from two scenes, N3DV~\cite{li2022neural} (Coffee Martini) and Meet Room~\cite{li2022streaming} (Discussion). As shown in Fig.\ref{fig:framelevel_N3DV} and Fig.\ref{fig:framelevel_meetroom}, across these sampled frames, our proposed StreamLoD-GS consistently produces higher-quality reconstructions. Compared to the pronounced artifacts in 3DGStream~\cite{sun20243dgstream}, the loss of fine-scale details in QUEEN~\cite{girish2024queen}, and the temporal overfitting observed in 4DGC~\cite{hu20254dgc}, our approach delivers more stable results with richer and more accurate scene details.

\begin{figure}[!htbp]
\centering
\includegraphics[width=0.45\textwidth]{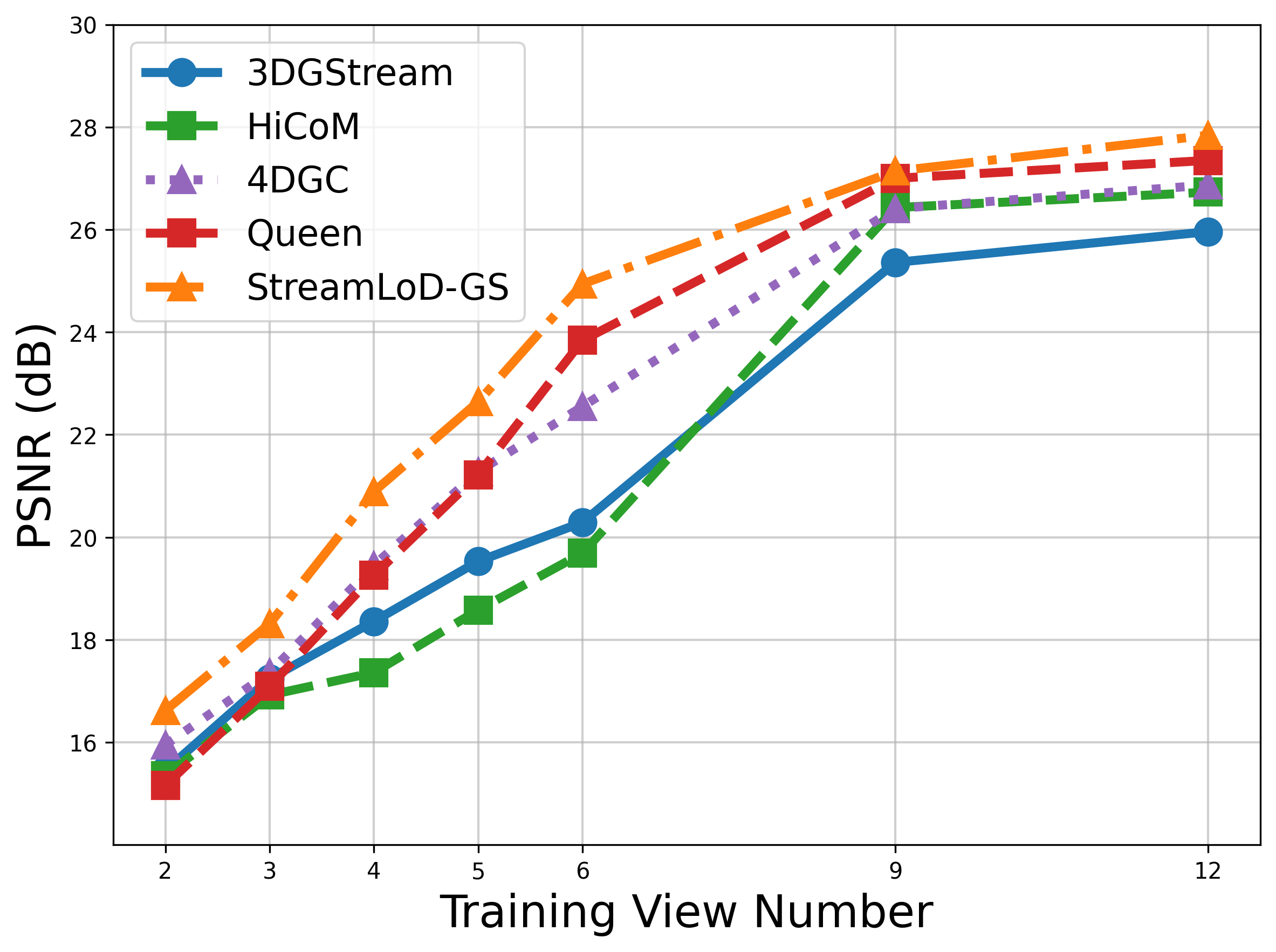} 
\caption{Quantitative comparison in Meet Room~\cite{li2022streaming} dataset with varying numbers of training views. }
\label{fig:Sparseview_performance}
\end{figure}  
     
\begin{table}[!htbp]
\centering
\caption{Quantitative Comparison on the Meet Room~\cite{li2022streaming} dataset with 3 training views. HD-GS: Hierarchical Gaussian Dropout. $\text{F}_{\text{seps}}$ indicates a variant that replaces the unified attribute-prediction MLP ($\text{F}_{\text{mlp}}$) with the separate networks for estimating Gaussian attributes.}
\begin{tabular} {l|c|c|c|c}
    \hline
    \multirow{2}{*}{\textbf{Methods}} & PSNR & Storage & Train & Render \\
    & (dB)$\uparrow$ & (MB)$\downarrow$ & (s)$\downarrow$ & (FPS)$\uparrow$ \\
    \hline
     w $\text{F}_{\text{seps}}$ & 18.68 & 0.290 & 1.330 & 425  \\
     w/o HD-GS & 17.99 & 0.256 & 1.096 & 498  \\
     \hline
   LoD-AO  & 18.85 &0.253 & 1.073 & 547  \\
    \hline
\end{tabular} 
\label{table:ablation_sep}
\end{table}

\begin{figure*}[!htbp]
\centering
\includegraphics[width=0.99\textwidth]{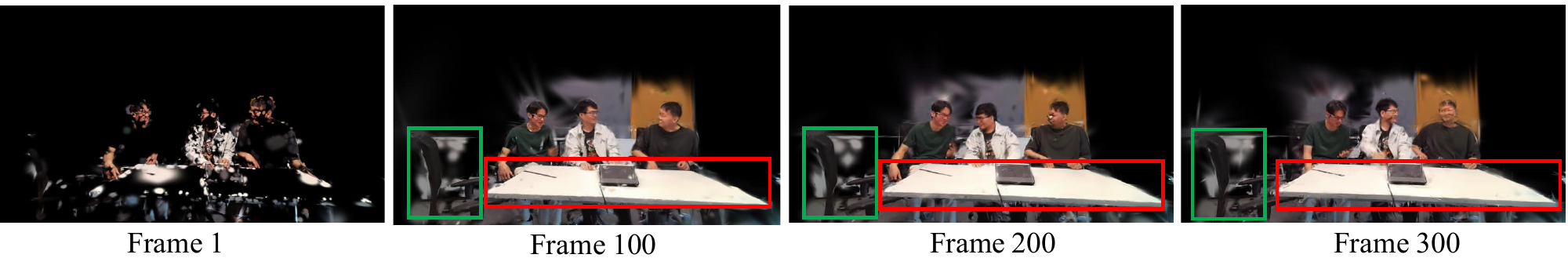} 
\caption{Rendering results with dynamic anchors, which are separated by using our GMM-based motion partitioning strategy, on a set of equally spaced frames from the Meet Room (Scene: Discussion) dataset~\cite{li2022streaming}.}
\label{fig:limitation}
\end{figure*}

\subsection{Ablation Study}
\subsubsection{Analysis of Module Contributions}
We evaluate the individual contributions of three core components: LoD-Structured 3DGS with Anchor and Octree (LoD-AO), GMM-Based Motion Partitioning (GMM-Part), and Quantized Residual Refinement (Q).
As summarized in Tab.~\ref{table:ablation}, the full configuration of StreamLoD-GS achieves the optimal balance across all metrics, including SSIM, LPIPS, storage efficiency, and rendering throughput.
Specifically, removing the Quantized Residual Refinement module leads to a marginal decline in reconstruction quality while inflating the storage footprint to 1.683 MB. The exclusion of GMM-Part results in increased storage and training latency, with rendering speed dropping to 425 FPS.
Notably, omitting LoD-AO reduces training time to 0.902 s, but at the cost of a severe 2.71 dB PSNR drop (from 22.73 dB to 20.02 dB) and a more than twofold increase in storage. These results underscore that each component is indispensable and that they work synergistically to maintain high performance.

\subsubsection{Analysis of LoD-AO with Hierarchical Dropout and Shared MLP }
Within the LoD-AO framework, we introduce Hierarchical Gaussian Dropout (HD-GS)—a level-aware strategy designed to mitigate overfitting under Sparse-View  constraints and stabilize the optimization process. Furthermore, to avoid the parameter explosion associated with multiple networks, we employ a shared MLP ($\text{F}_{\text{mlp}}$) to predict attributes for hierarchical Gaussians. Unlike Scaffold-GS~\cite{lu2024scaffold}, which utilizes separate networks for this task, our shared architecture accelerates both training and rendering without compromising fidelity.
To validate these design choices, we compare our shared $\text{F}_{\text{mlp}}$ against a variant using separate networks ($\text{F}_{\text{seps}}$) and evaluate the impact of removing the HD-GS module. As shown in Tab.~\ref{table:ablation_sep}, the $\text{F}_{\text{seps}}$ variant exhibits a noticeable decrease in PSNR and slower rendering throughput, confirming the efficiency and efficacy of the shared network. Most significantly, disabling the HD-GS module results in the most substantial decline in PSNR, demonstrating that hierarchical dropout is critical for maintaining reconstruction fidelity and enabling robust optimization in Sparse-View settings.
These findings highlight that the integration of $F_{\text{mlp}}$ and HD-GS allows LoD-AO to achieve peak rendering speeds (547 FPS) and minimal storage costs while maximizing training efficiency.

\subsubsection{Analysis of Different Training Views}
Fig.~\ref{fig:Sparseview_performance} illustrates the reconstruction performance of various methods on the Meet Room~\cite{li2022streaming} dataset across a range of training view counts (2, 3, 4, 5, 6, 9, and 12). For each configuration, all available views are reserved for testing to ensure a rigorous assessment of novel-view synthesis. 
As depicted in the results, all evaluated methods exhibit a positive correlation between PSNR and the number of training views, confirming that additional viewpoints provide essential geometric and photometric constraints for scene reconstruction.
Notably, StreamLoD-GS consistently outperforms 3DGStream~\cite{sun20243dgstream}, HiCoM~\cite{gao2024hicom}, 4DGC~\cite{hu20254dgc}, and QUEEN~\cite{girish2024queen} across all view densities. This sustained performance gap underscores the superior robustness and generalization of our framework under Sparse-View constraints. These findings highlight StreamLoD-GS's exceptional ability to effectively reconstruct high-quality dynamic scenes even when provided with severely limited training data.
 
\subsection{Limitation} 
As illustrated in Fig.~\ref{fig:limitation}, our GMM-based motion partitioning strategy effectively decouples static and dynamic regions as the video sequence progresses. However, a notable limitation persists when processing static, visually homogeneous surfaces with specular reflections, such as the reflective chair surfaces, the central laptop, and the uniform white desk (indicated by the red and green boxes in Fig.~\ref{fig:limitation}). Specifically, the view-dependent appearance changes induced by reflections introduce temporal inconsistencies, a well-recognized challenge in recent literature~\cite{yao2025stdd, liu2024temporally}. These subtle photometric variations across consecutive frames generate phantom gradients that can lead the GMM to misclassify static regions as dynamic. 
Consequently, enhancing the robustness of motion segmentation for visually uniform areas, particularly under complex lighting and reflective conditions, remains a critical objective. In future work, we aim to address this by integrating both color-consistency constraints and 3D geometric priors to better distinguish true motion from specular artifacts, thereby improving the precision of dynamic anchor identification.

\section{Conclusion}
\label{sec:Conclusion}
In this work, we introduce StreamLoD-GS, a novel hierarchical Level-of-Detail (LoD) Gaussian Splatting framework designed for efficient streaming Free-Viewpoint Video (SFVV) reconstruction. To mitigate overfitting and stabilize the optimization process, we incorporate a Hierarchical Gaussian Dropout technique, which selectively drops unnecessary Gaussians.
StreamLoD-GS also features a GMM-based motion partitioning module that separates the 3DGS into dynamic and static regions, represented by dynamic and static anchors, respectively. During subsequent training, static anchors remain frozen, while dynamic anchors are updated in a streaming manner as new frames arrive.
Additionally, a Quantized Residual Refinement module is introduced to quantize the residuals of the dynamic anchors, effectively reducing storage costs without compromising performance.
To assess the effectiveness of StreamLoD-GS, we conducted extensive experiments across various view settings. The results show that our method outperforms existing approaches, delivering significant improvements in storage efficiency, rendering speed, and visual quality.


\bibliographystyle{IEEEtran}
\bibliography{main}

\newpage
\vspace{11pt}
\vspace{-33pt}
\begin{IEEEbiography}[{\includegraphics[width=1in,height=1.25in,clip,keepaspectratio]{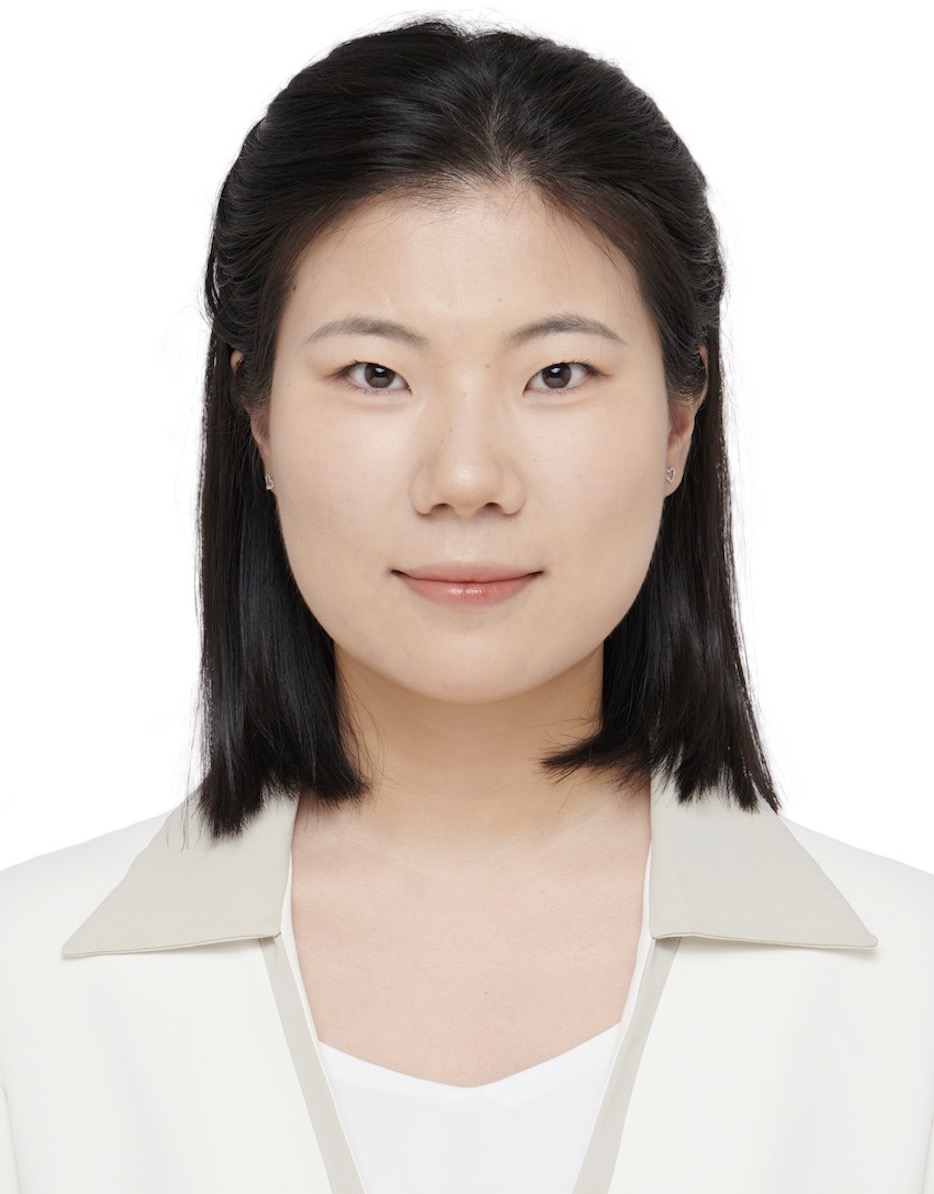}}]{Xinhui Liu} received the B.S. and M.S. degrees in electronic information engineering from Xi’an Polytechnic University, Xi’an, China, in 2015 and 2018, respectively. She received the Ph.D. degree from Xi’an Jiaotong University, Xi’an, China, in 2024. From Sep. 2022 to May. 2024, she was a visiting student in the University of Sydney, NSW, Australia, and work with Prof. Luping Zhou. She is currently a Postdoctoral Researcher with the School of Computing and Data Science, the University of Hong Kong, Hong Kong, China. Her current research interests include computer vision and Computer Graphics.
\end{IEEEbiography}
\vspace{11pt}
\vspace{-33pt}
\begin{IEEEbiography}[{\includegraphics[width=1in,height=1.25in,clip,keepaspectratio]{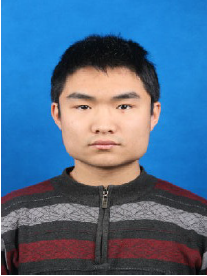}}]{Can Wang} 
is currently a research fellow at The University of Hong Kong.
He received the Ph.D. degree in the Department of Computer Science, City University of Hong Kong, HK, People’s Republic of China. He got his M.S. in Computer Science and Technology from the University of Science and Technology of China in 2020. His current research interests include computer graphics and computer vision.
\end{IEEEbiography}
\vspace{11pt}
\vspace{-33pt}
\begin{IEEEbiography}[{\includegraphics[width=1in,height=1.1in,clip,keepaspectratio]{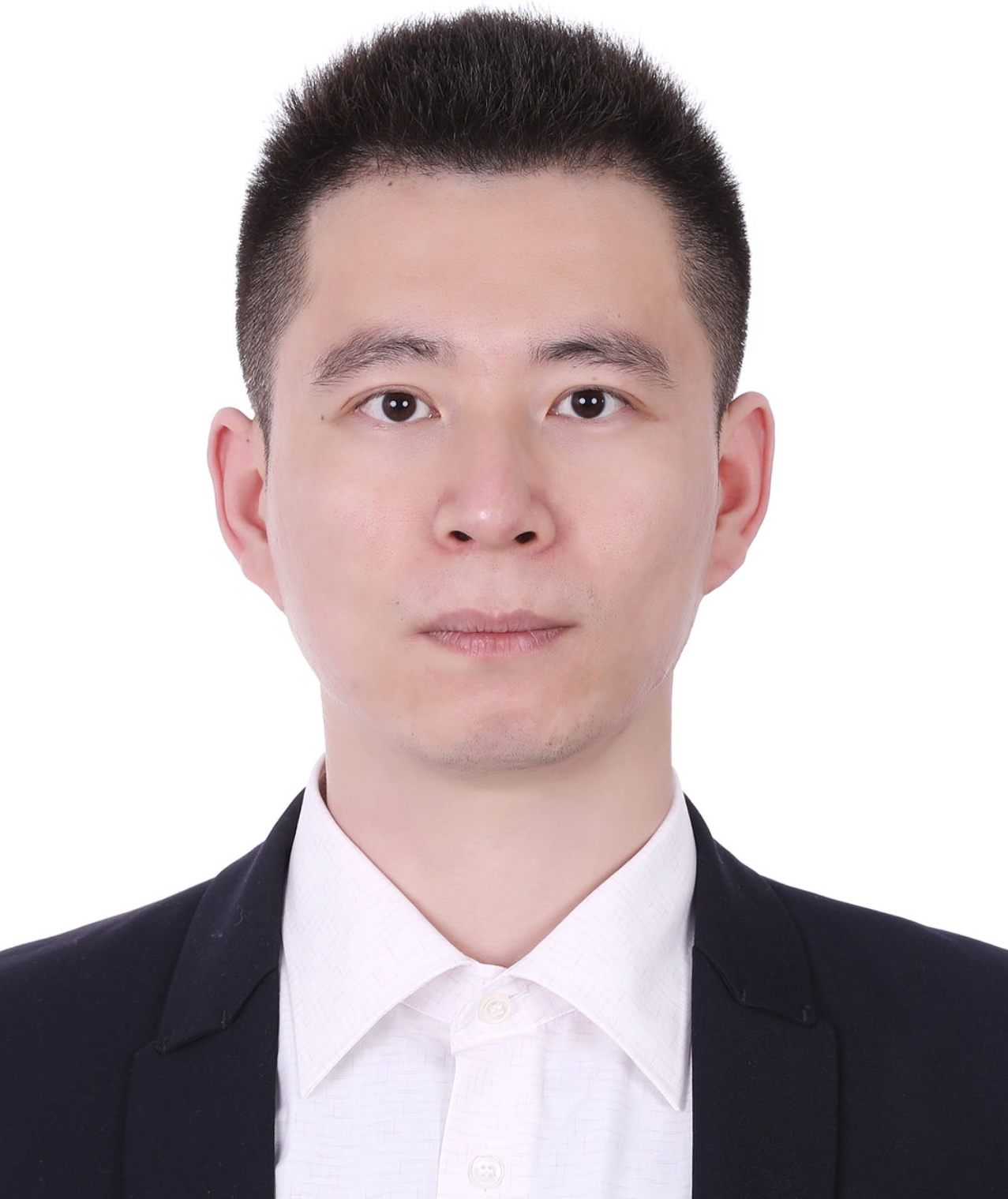}}]{Lei Liu} Lei Liu is a Postdoctoral Fellow at the University of Hong Kong. He received his Ph.D. degree from Beihang University, China, in 2025. He obtained his Bachelor's and Master's degrees from the College of Instrumentation and Electrical Engineering, Jilin University, in 2017 and 2020, respectively. His research interests include visual data compression and 3D Gaussian Splatting editing.
\end{IEEEbiography}
\vspace{11pt}
\vspace{-33pt}
\begin{IEEEbiography}[{\includegraphics[width=1in,height=1.1in,clip,keepaspectratio]{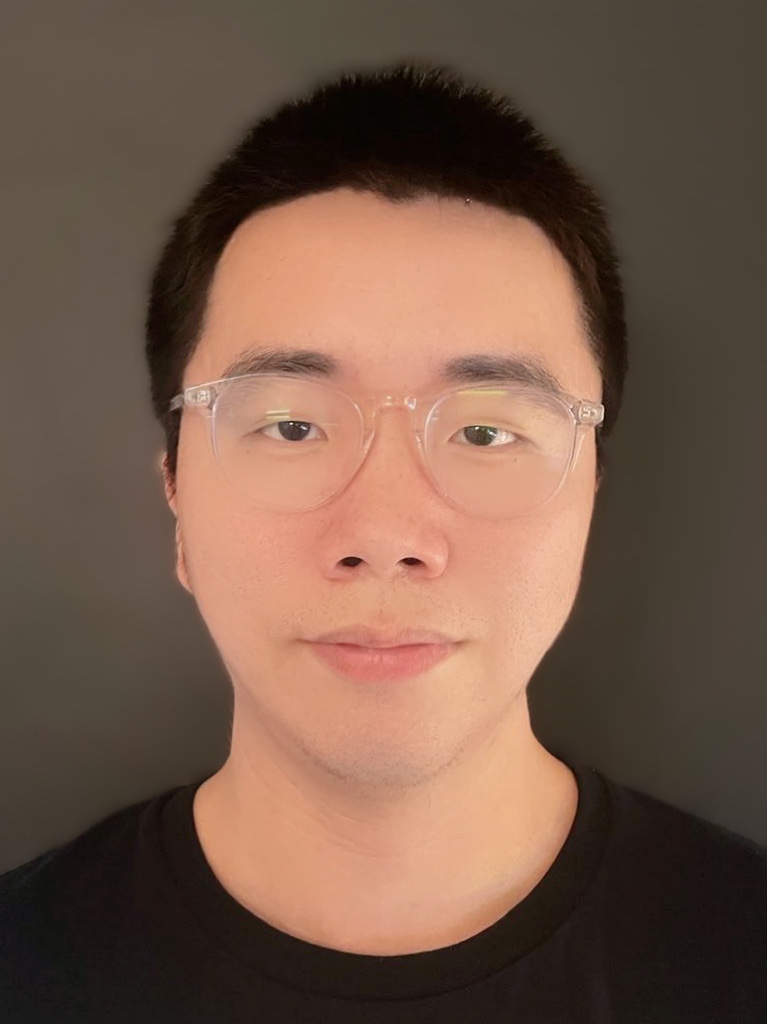}}]{Zhenghao Chen} Dr. Zhenghao Chen is an Assistant Professor at the University of Newcastle. He obtained his B.IT. H1 and Ph.D. from the University of Sydney, in 2017 and 2022. Previously, he was a Research Engineer at TikTok, a Research Fellow at the University of Sydney, and a Visiting Research Scientist at Microsoft Research and Disney Research. Dr. Chen's general research interests encompass Computer Vision and Machine Learning.
\end{IEEEbiography}
\vspace{11pt}
\vspace{-33pt}
\begin{IEEEbiography}[{\includegraphics[width=1.1in,height=1.25in,clip,keepaspectratio]{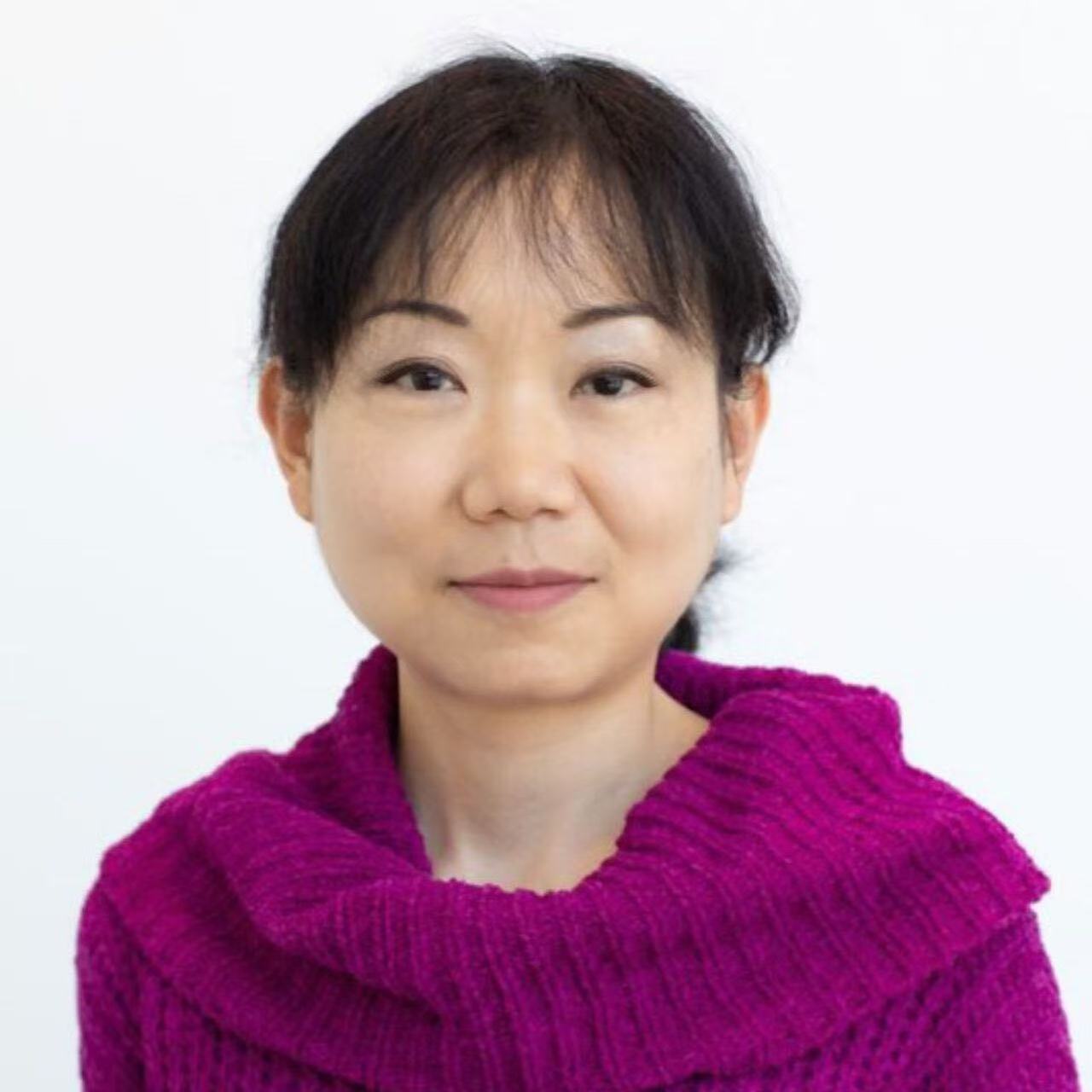}}]{Wei Jiang } received her PhD degree in E.E. from Columbia University in 2010, her M.S. and B.E. degrees in Automation from Tsinghua University in 2005 and 2002, respectively. She has published over 40 refereed papers and owned over 40 issued patents in the field of computer vision and machine learning, and holds 10+ standard adoptions in IEEE and JVET/MPEG/JPEG standards. She is a senior member of IEEE and serves as TPC member and reviewers for several top conferences and journals. Wei Jiang is currently a Senior Principal Researcher in Futurewei Technologies. She has broad research interests in computer vision and artificial intelligence, including AI-based image and video compression, image and video restoration and generation, visual computing, and multimedia content analysis. Her current research is focused on research and standardization of next-generation AI-based image/video compression and restoration, image/video generation by vision-language models, and inference acceleration.
\end{IEEEbiography}
\vspace{11pt}
\vspace{-33pt}
\begin{IEEEbiography}[{\includegraphics[width=1.2in,height=1.25in,clip,keepaspectratio]{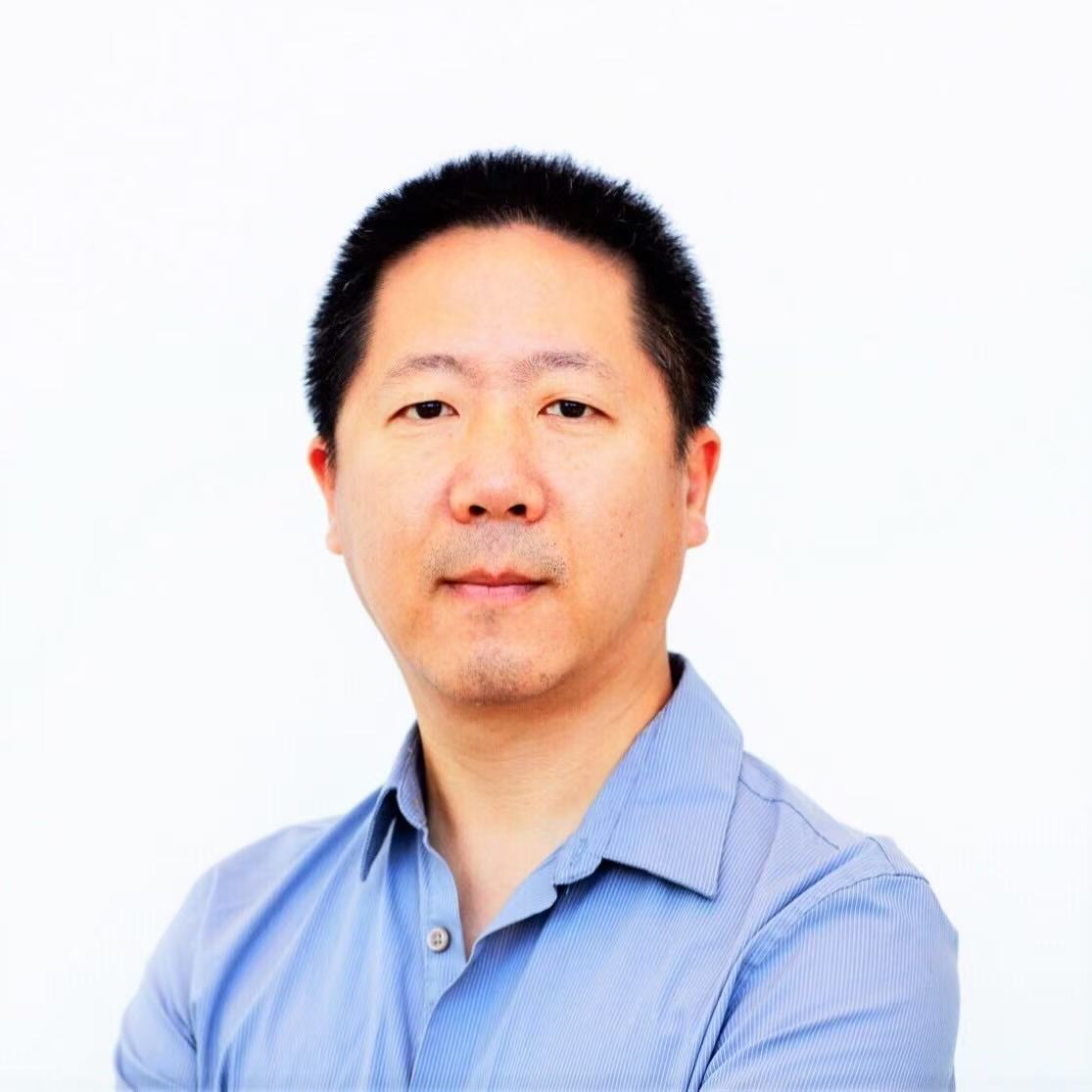}}]{Wei Wang} received his M.S. and B.S. degree from E.E. Department of Fudan University, China, in 1998 and 1995, respectively. He is currently a Principal Researcher in Futurewei Technologies, He was formerly a Senior Staff Researcher in Alibaba Group.
He has been involved in international standardization activities, including contributing to ISO/IEC Moving Picture Experts Group for work items on H.265/HEVC SCC and neural network representation, and to IEEE Data Compression Standards Committee for work item on neural image coding. He served as co-Chair of MPEG neural network representation group. 
His current research interests include high performance computing, image and video compression, neural networks compression and acceleration, and converting algorithms to product on desktop, embedded or ASIC platforms. 
\end{IEEEbiography}
\vspace{11pt}
\vspace{-33pt}
\begin{IEEEbiography}[{\includegraphics[width=1in,height=1.25in,clip,keepaspectratio]{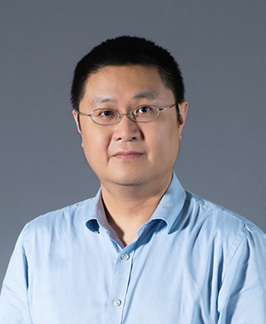}}]{Dong Xu (Fellow, IEEE)} received the B.E. and Ph.D. degrees from the University of Science and Technology of China, Hefei, China, in 2001 and 2005, respectively. While pursuing the Ph.D. degree, he was an Intern with Microsoft Research Asia, Beijing, China, and a Research Assistant with The Chinese University of Hong Kong, Hong Kong, for more than two years. He was a Postdoctoral Research Scientist at Columbia University, New York, NY, USA, for one year. He also worked as a Faculty Member at Nanyang Technological University, Singapore, and the Chair of computer engineering at The University of Sydney, NSW, Australia. He is currently a Professor with the School of Computing and Data Science, The University of Hong Kong, Hong Kong, China.
He was the co-author of a paper that received the Best Student Paper Award from the IEEE Conference on Computer Vision and Pattern Recognition (CVPR) in 2010 and a paper that received the Prize Paper Award in IEEE Transactions on Multimedia in 2014. His current research interests include Artificial Intelligence, Computer Vision, Multimedia and Machine Learning. 
\end{IEEEbiography} 
\vfill
\end{document}